# Revenue-based Attribution Modeling for Online Advertising

Kaifeng Zhao
Data and Analytics R&D, GroupM, Singapore
Kaifeng.zhao@groupm.com

Seyed Hanif Mahboobi
Data and Analytics R&D, GroupM, New York, NY, USA
Seyed.Mahboobi@groupm.com

Saeed R. Bagheri
Data and Analytics R&D, GroupM New York, NY, USA
Saeed.Bagheri@groupm.com

# Revenue-based Attribution Modeling for Online Advertising

## Abstract

This paper examines and proposes several attribution modeling methods that quantify how revenue should be attributed to online advertising inputs. We adopt and further develop relative importance method, which is based on regression models that have been extensively studied and utilized to investigate the relationship between advertising efforts and market reaction (revenue). Relative importance method aims at decomposing and allocating marginal contributions to the coefficient of determination ($R^2$) of regression models as attribution values. In particular, we adopt two alternative submethods to perform this decomposition: dominance analysis and relative weight analysis. Moreover, we demonstrate an extension of the decomposition methods from standard linear model to additive model. We claim that our new approaches are more flexible and accurate in modeling the underlying relationship and calculating the attribution values. We use simulation examples to demonstrate the superior performance of our new approaches over traditional methods. We further illustrate the value of our proposed approaches using a real advertising campaign data set.

# 1 Introduction

Digital advertising revenue in the U.S. is reported to be $27.45bn for the first half of 2015 and has experienced a compound annual growth rate of 17% from 2005 to 2014 (IAB, 2014). Digital advertising exposes marketers to the full tracks of users' conversion paths (i.e. sequence and timing of, as well as engagement in advertising channels that have reached them) in addition to their demographic information, shopping preferences and many other relevant information. By analyzing this data, marketers have the potential to gain better insights into how advertisements have an impact on the users and consequently effect their purchasing activity. These insights further help to make smarter decisions in the planning of investments in advertising.

A variety of online channels, including search, display, digital video, etc., are employed in digital advertising campaigns. Each individual user is usually exposed to a combination of channels before making any purchasing decision. Therefore, a fundamental problem in measuring the advertising effectiveness and efficiency is to find out how revenue should be attributed to a group of advertising channels. The attribution problem studied in this paper can be formulated as follows. Given a sequence of revenue records ($y$) and users' exposure history to $p$ advertising channels ($x_1, \ldots x_p$), we quantify how $y$ should be attributed to $x_j$'s and report attribution values $\phi_{x_j}$ for $j = 1, \ldots, p$. We use this notation consistently for all of the methods discussed throughout this paper.

Attribution models and methods offer systematic approaches to solve the attribution problem. A common attribution method used by industry practitioners is the last-touch (first-touch) model. It assigns all credits to the last (first) channel a converted user has been exposed to. However, this approach is flawed because it systematically undervalues channels that expose customers farther from (closer to) the conversion funnel.

To better allocate individual contributions and synergies, algorithm-based methodologies have increasingly attracted attention over the past decade (Abhishek et al., 2012; Anderl et al., 2013; Dalessandro et al., 2012; Li and Kannan, 2013; Shao and Li, 2011; Tucker, 2012; Wiesel et al., 2011). Shao and Li (2011) have proposed a "simple probabilistic model". It examines, in addition to each individual channel's effect, the joint effect of all possible combinations of 2, 3 or more channels. Subsequently, Dalessandro et al. (2012) have developed a channel importance attribution method as a generalization of Shao and Li (2011). This method analyzes channel interactions up to full order, the so-called joint effects of all advertising channels. The resulting approach, as discussed in their work, is equivalent to the Shapley Value method in Cooperative Game Theory (Osborne and Rubinstein, 1994; Shapley, 1952). These approaches are classified as direct methods as they directly analyze the channels' contribution in terms of credits (conversion, conversion probability or revenue). On a parallel line of development, regression models have become very popular in investigating the relationship of revenue and advertising efforts (Cain, 2010). Attribution methods involving an auxiliary model like regression are considered as indirect attribution approaches because the attributions are based on a presumed (regression) model of the relationship between advertising exposure and revenue. To the best of our knowledge, we find relative importance methods as the only set of relevant regression-based approaches for the revenue-based attribution problem among the vast statistical literature on regression models.

Relative importance methods aim at decomposing the coefficient of determination ($R^2$). This decomposition is performed either by dominance analysis or relative weight analysis. Dominance analysis (Azen and Budescu, 2003; Budescu, 1993; Tonidandel and LeBreton, 2011) examines all nested submodels, i.e., models involving subsets of variables. It derives from the Shapley Value concept in Cooperative Game Theory (Osborne and Rubinstein, 1994; Shapley, 1952) to measure the marginal $R^2$ by adding each variable to every possible submodel. Attribution values are then calculated as aggregations of all marginal effects. This analysis suffers from computational efficiency issues; the number of submodels to be fitted grows exponentially in the number of channels. Relative weight analysis is an alternative approach developed by Johnson (2000). This analysis generates a set of auxiliary uncorrelated variables based on singular value decomposition. It then calculates attribution values for those auxiliary variables that are consequently transformed back to the originals. This analysis is shown to produce very similar results to those produced by dominance analysis at a substantially higher computational efficiency (Tonidandel and LeBreton, 2011).

In this paper, we implement several revenue-based multi-touch attribution approaches and further develop relative importance method. We extend the relative importance method from linear regressions to additive models: a semi-parametric model type. A linear regression model implicitly assumes that the advertising channels contribute to revenue in a linear manner. As it is generally a restrictive assumption to expect the underlying data generation process of any real dataset to be linear, the resulting attribution values are very likely to be inaccurate. Nonlinear parametric models, such as logistic regression, might be viable alternatives. Nonetheless, it is still difficult to justify postulating any pre-assumed model structure (Yadagiri et al., 2015). Synergies among advertising channels can be very high and varying largely with country, user segment, industry, product and many other factors and factor combinations. In other words, a good model should allow for far greater flexibility in unfolding the underlying relationship. Additive model, as a semi-parametric model, is popularly used in the fields of econometrics, statistics and machine learning (Friedman et al., 2001; Hastie and Tibshirani, 1990; Murphy, 2012) to achieve a degree of flexibility. While it still incorporates a parametric component which controls the overall model, a non-parametric component provides flexibility. As such, semi-parametric approaches have the potential to be an important auxiliary modeling paradigm for indirect attribution approaches.

The rest of the paper is organized as follows. In Section 2, we review and discuss several conventional attribution approaches on a linear regression model. We then present a detailed description of our proposed relative importance methods in Section 3. Section 4 demonstrates the performance of the proposed methodologies using both simulated and real data examples. We conclude our work with a brief discussion in Section 5.

## 2 Review of Legacy Methods

Various methods have been developed to measure the importance of variables prior to the advent of relative importance methods. They are mostly based on correlations among variables and coefficients from a standard linear model of $y$ and $\boldsymbol{X} = (x_1, \ldots x_p)$ defined as following,

$$y = \beta_1 x_1 + \cdots + \beta_j x_j + \cdots + \beta_p x_p + \varepsilon, \tag{1}$$

where $\varepsilon$ is an error term that follows the same normal distribution with mean 0 and variance $\sigma^2$, independently across all observations. Without loss of generality, we presume that the dependent variable $y$ and the variables $x_j$'s are all standardized to have a mean of 0 and variance of 1 so that the intercept can be excluded from the analysis. We denote the coefficient of determination of this model as $R^2_{y,X}$.

### 2.1 Regression Coefficients

Regression coefficients are perhaps the most conventional measures used for assessing variable importance. The vector $\boldsymbol{\beta} = (\beta_1, \ldots, \beta_p)$ represents the changes in $y$ associated with unit change in each independent variable with others left unchanged. In case of no inter-correlations among variables, $\beta_j$ is equivalent to the correlation of $y$ and $x_j$, which is denoted as $\rho_{y,x_j} = cor(y, x_j)$. Moreover, we have

$$R^2_{y,X} = \sum_{j=1}^{p} \rho^2_{y,x_j} = \sum_{j=1}^{p} \beta_j^2. \tag{2}$$

This equation implies that the squared coefficients ($\beta_j^2$'s) perfectly partition the $R^2_{y,X}$ under no inter-correlations among variables and thus are effective measures of variable importance. We find three naïve bases of attribution that uses this fact: $\beta_j^2$, $\rho^2_{y,x_j}$, and $\beta_{x_j} \rho_{y,x_j}$, $j = 1, \ldots, p$. As noticed in a line of literature (Budescu, 1993; Green and Tull, 1974; Hoffman, 1960), equation (2) can have substantial deviations in case of non-zero inter-correlation among variables. Therefore, using regression coefficients as bases of attribution is inappropriate because it ignores the interactions among advertising channels.

### 2.2 Squared Correlations

Squared correlation $\boldsymbol{\rho}^2 = (\rho^2_{y,x_1}, \ldots, \rho^2_{y,x_p})$, as mentioned in 2.1, is another obvious and popular measure of variable importance. However, it should be even less preferred in case of non-zero inter-correlation. This is because it only utilizes the correlation of $y$ and each individual $x_j$, and consequently ignores the overall relationship of variables as captured by the regression model.

## 2.3 Product of Regression Coefficients and Correlations

Having $\boldsymbol{\beta}$ and $\boldsymbol{\rho}$ defined in the previous sections, one may naturally combine them and consider the product, $\beta_{x_j}\rho_{y,x_j}$, $j = 1, \ldots, p$ (Azen and Budescu, 2003; Tonidandel and LeBreton, 2011). This method becomes difficult to justify and implement especially when the product is negative.

# 3 Relative Importance Methods

Relative importance methods aim at decomposing the coefficient of determination ($R^2$) of regression models. Basically, we need to 1) choose/fit a regression model that describes the relationship of revenue and advertising efforts; 2) decompose the resulting $R^2$. In this section, we present a detailed description of proposed approaches, including $R^2$ decomposition methods (dominance analysis and relative weight analysis) and how we implement them to our choices of regression models (linear and additive models).

## 3.1 $R^2$ Decomposition

As an alternative to using regression coefficient and correlation as a proxy of changes in the coefficient of determination ($R^2$), relative importance methods decompose and allocate $R^2$ directly. The $R^2$ of a regression model represents the portion of the dependent variable variance that can be explained by the fitted model to a proper subset of variables. The coefficient of determination cannot decrease when the subset adds new variables. Therefore, the resulting attribution values are always non-negative. In this section, we introduce dominance analysis and relative weight analysis on the standard linear model as expressed in (1).

### 3.1.1 Dominance Analysis

Dominance analysis (DA) (Azen and Budescu, 2003; Budescu, 1993) compares coefficients of determination of all nested submodels composed of subsets of independent variables (one covariate out, tuple erasures, triple erasures, groups of variables, all possible choices, etc.) with that of the full model. Evaluating all submodels ensure interactions are fully taken into account for calculating attribution values. More precisely, DA proceeds as follows:

- Calculate $R^2_{y,x}$ for each submodel (out of a total of $2^p - 1$ submodels). Here, we use $R^2_{y,x_j,h}$ to denote the explained variance by submodel that contains variables with indices in $\{j\} \cup h$, where $h$ is any subset of $\{1, \ldots, p\}\setminus\{j\}$.
- Compare pairwise relative importance for each pair of variables (totally $p(p-1)/2$ such pairs). Specifically, we compare $R^2_{y,x_i,h}$ and $R^2_{y,x_j,h}$ with $i \neq j$ where $h \subseteq \{1, \ldots, p\}\setminus\{i,j\}$.
- $C^k_{x_j}$ measures the marginal contribution of variable $x_j$ when added to a submodel consisting of $k$ variables excluding $x_j$. There are $\binom{p-1}{k}$ such submodels. $C^k_{x_j}$ is calculated by averaging the associated incremental $R^2$,

$$C_{x_j}^k = \sum_{|h|=k} (R_{y,x_j,h}^2 - R_{y,\emptyset,h}^2) / \binom{p-1}{k} \tag{3}$$

- Attribution value of $x_j$ is therefore defined as,

$$\phi_{x_j} = \sum_{k=0}^{p-1} C_{x_j}^k / p. \tag{4}$$

- Attribution values partition $R_{y,X}^2$:

$$R_{y,X}^2 = \sum_{j=1}^{p} \phi_{x_j}. \tag{5}$$

We finally define two concepts of covariate dominance:

— Complete dominance
 $x_i$ completely dominates $x_j$ if $R_{y,x_i,h}^2 \geq R_{y,x_j,h}^2$ for all $h \subseteq \{1, \ldots, p\} \setminus \{i, j\}$.

— General dominance
 $x_i$ generally dominates $x_j$ if $\phi_{x_i} \geq \phi_{x_j}$.

Dominance concepts can powerfully draw a complete picture for the patterns of variables' relative importance. Through dominance analysis, one is able to address the problems such as whether a particular (group of) variable(s) is (are) important than another in presence or absence of some other variable(s). An inherent problem with dominance analysis is computational efficiency: there are $2^p - 1$ submodels to be estimated in order to obtain attribution values.

### 3.1.2 Relative Weights Analysis

As mentioned in 2.1, squared regression coefficient $\beta_j^2$ is a flawed measure for variable importance as it ignores inter-correlations. On the other hand, exact attributions are very difficult to obtain using dominance analysis. The idea of relative weight (RW) analysis is to create a new set of orthogonal variables from the original ones to get rid of inter-correlations. Therefore, $\beta_j^2$'s of these auxiliary variables become directly usable as relative importance values. Moreover, because these variables are linear combinations of the original variables, they can be easily converted back to the original ones.

We suppose that the independent variable matrix consisting of $n$ observations, $X \in \mathbb{R}^{n \times p}$ can be decomposed by using singular value decomposition as

$$X = P\Delta Q', \tag{6}$$

where columns of $P$ are the eigenvectors of $XX'$. Columns of $Q$ contain the eigenvectors of $X'X$. $\Delta$ is a diagonal matrix containing the singular values of $X$. Singular values are the square roots of the eigenvalues of $X'X$ and $XX'$. Therefore, a best-fitting orthogonal approximation of $X$ (i.e., an approximation $Z$ which minimizes the elementwise sum of squares of $X - Z$) can be given as,

$$Z = PQ', \qquad (7)$$

in which columns of $Z$ are orthogonal vectors that are not correlated with each other. The vector of coefficients when regressing $y$ on the orthogonal variables $Z$ is obtained as

$$\beta^* = (Z'Z)^{-1}Z'y = QP'y. \qquad (8)$$

Because columns of $Z$ are uncorrelated, $\beta^{*2}$ are effective measures of the relative importance of $Z$ variables on $y$. Next, we regress the columns of $X$ on $Z$ to retrieve the importance values of $X$.

$$X = Z\Lambda, \text{ with } \Lambda = (Z'Z)^{-1}Z'X = Q\Delta Q', \qquad (9)$$

where $\Lambda$ is the matrix of regression weights of $Z$ on $X$. By realizing that $Z$ is simply a linear transformation of $X$, relations between $\beta^{*2}$ and the relative weights of $X$ variables on $y$ can be recovered as

$$\phi = \Lambda^2 \beta^{*2}. \qquad (10)$$

Finally, we have $\sum_j \phi_{x_j} = R^2_{y,X}$ with $\phi_{x_j}$ the $j$-th element of vector $\phi$.

An immediate benefit from relative weights analysis is computational efficiency, since there is no need to examine multiple submodels. Also, relative weight analysis and dominance analysis generally produce quite similar relative importance values (Tonidandel and LeBreton, 2011).

## 3.2 Regression Models

To calculate the attribution values in our problem, the $R^2$ decomposition should be performed on regression models that capture the relationship of revenue and advertising inputs. However, to the best of our knowledge, all existing $R^2$ decomposition methods are developed for standard linear models. With parametric restrictions on model structure, a linear regression may fail to accurately describe the true underlying relative importance. In order to obtain better modeling capability and more flexibility, we enrich our choices of regression models in relative importance methods by extending the $R^2$ decomposition to a semi-parametric model.

### 3.2.1 Linear Models

As previously discussed, the regression model, as specified in (1), on which the $R^2$ decomposition methods are based, is the simplest variation of linear models. However, the linear model also refers to a broader class of regression models that is linear in coefficients. In other words, the nonlinear transformation of independent variables (such as quadratic, cubic, etc.) is allowed. The resulting regression model is still considered to be linear as long as the dependent variable is represented as a linear combination of other (transformed) variables. In this sense, it is also worth mentioning that the $R^2$ decomposition can be easily extended to those models by treating the transformed variables in the same way as their untransformed counterparts.

### 3.2.2 Additive Models

In contrast to a standard linear model which assumes model components arise from known parametric forms, additive model is known to be more flexible for incorporating nonparametric model components (Friedman et al., 2001; James et al., 2013). It has the following general form

$$\boldsymbol{Y} = \sum_{j=1}^{p} f_j(\boldsymbol{x}_j) + \varepsilon, \tag{11}$$

where $f_j(\boldsymbol{x}_j)$'s are unknown functions that operate on variables $\boldsymbol{x}_j = (x_{1,j}, \dots, x_{n,j})$ with $n$ as the number of observations, and $\varepsilon$ as the error term. Note that no functional form is pre-specified for $f_j$'s which are left to be estimated. In what follows, we estimate the additive model by rewriting it into a linear model using a truncated power spline (TPS), and then adopt the relative importance methods for it.

TPS is a popular and easy approximation scheme to estimate non-parametric functions (James et al., 2013; R. and de Boor, 1980; Zhao and Lian, 2014). Under certain continuity and smoothness conditions, it can be used to approximate functions with high accuracy and computational efficiency. Specifically, TPS approximates the unknown function $f_j$ by using a linear combination of the following bases (denoted with $B_0(x), B_1(x), \dots, B_{q+K}(x)$),

$$1, x, x^2, \dots, x^q, (x - t_1)^q \, I\{x > t_1\}, \dots, (x - t_K)^q \, I\{x > t_K\} \tag{12}$$

where $q$ is the highest order of polynomial and is commonly set to be 3 (cubic spline), and $t_0 = \min_i(x_i) < t_1 < t_2, \dots, < t_K < \max_i(x_i) = t_{K+1}$ partition $[\min_i(x_i), \max_i(x_i)]$ into subintervals $[t_k, t_{k+1}), k = 0, \dots, K$, with $K$ internal knots. $I\{\cdot\}$ is the indicator function: it takes the value 1 if the condition within the bracelets hold; it is 0 otherwise. In practice, $K$ is a tuning parameter, and is tuned using cross validation.

We choose to use equally spaced knots. For example, if $\max_i(x_i) = 1$ and $\min_i(x_i) = 0$, the equally spaced knots with $K = 4$ would be 0.2, 0.4, 0.6 and 0.8.

Given well-defined functional bases, we proceed to estimate the following,

$$f_j(x_j) = \sum_{k=0}^{q+K} b_{j,k} B_{j,k}(x_j), j = 1, \ldots, p, \tag{13}$$

where $b_{j,k}, k = 0, \ldots, q+K$ are the coefficients to be estimated and $B_{j,k}(x_j), k = 0, \ldots, q+K$ denotes the TPS bases created using variable $x_j$. By substituting (13) into model (11), we again have a linear model as it is linear in the coefficients ($b_{j,k}$'s). Therefore, the model can be estimated using the method of least squares.

Finally, we adopt the relative importance methods as introduced in Section 3.1 as a complement to the additive model. We first calculate the relative importance values of all TPS bases created using each of the original variables ($x_j$'s), and denote them as $\phi_{B_{j,k}}$ with $j = 1, \ldots, p$ and $k = 0, \ldots, q+K$. To retrieve the relative importance of $x_j$, we choose the summation of all terms that depend on $x_j$, the relative importance values $\phi_{B_{j,k}}$ with $k = 0, \ldots, q+K$. That is,

$$\phi_{x_j} = \sum_{k=0}^{q+K} \phi_{B_{j,k}}, j = 1, \ldots, p. \tag{14}$$

## 4   Numerical Results

In this section, we use various simulated and real datasets to demonstrate the performance of the proposed relative importance methods. They include dominance analysis with linear & additive models and relative weight analysis with linear & additive models.

We first use three simulation examples to illustrate the merits of the proposed relative importance methods. In example 1, we generate a synthetic dataset from a linear model to justify the superior performance of the proposed methods over the conventional metrics. The correlations (synergies) among variables are then increased in example 2. This is used to demonstrate that the proposed methods can incorporate such information and produce reasonable attribution values. Lastly, in example 3, we generate another dataset from an additive model with some nonlinearities. Relative importance methods based on linear models are compared with those based on additive models. Numerical results show that non-parametric functions in additive models can approximate various saturating and non-saturating transformation functions that are quite popular in marketing practice. Attribution values will be calculated accordingly.

Based on a real campaign dataset, we carry out both group and all-channel analysis to illustrate the performance of the proposed approaches. We analyze how revenue should be attributed to different categories of channels (publisher vs. DSP vs. paid search), as well as to all the individual channels involved in the campaign.

## 4.1 Simulation Example 1: Relative Importance Methods

In example 1, we apply the proposed relative importance methods to a simulated dataset. One hundred data points, $(x_i, y_i)$ with $i = 1, \ldots, 100$, are generated from a standard linear model as specified in (1), with $\boldsymbol{\beta} = (3, -4.5, -0.5, 3, -4)$. The variables $x_{i,j}$ are generated from the standard normal distribution with correlation matrix given by $Cor(x_{j_1}, x_{j_2}) = (1/2)^{|j_1 - j_2|}$. We consider independent and identically distributed errors generated from a standard normal distribution. We replicate this experiment 30 times.

Table 1 summarizes the theoretical importance values of channels suggested by $\beta^2, \rho^2, \beta \cdot \rho$, as well as the estimated dominance analysis (DA) and relative weight (RW) values averaged over 30 replicates. Note that all importance values have been normalized to sum up to unity. As mentioned earlier, $\beta^2, \rho^2$ and $\beta \cdot \rho$ are inappropriate to serve as effective measures, which can be validated by examining the importance values of variables $x_3$ and $x_4$. As we can see, $\beta_3^2$ is the smallest squared coefficient (negligible), but $x_3$'s squared correlation to the dependent variable is the third largest, which indicates a reasonably strong relationship with $y$. Using $\beta_3^2$ alone would likely flaw the true relationship. While for $x_4$, we observe zero squared correlation, but a reasonably large $\beta_4^2$ which shows a relatively large effect in predicting $y$. These issues are resolved in both DA and RW. They somehow combine the information conveyed by both $\beta$ and $\rho$ as well as the interactions among variables, resulting in relatively small but non-negligible importance values for $x_3$ and $x_4$.

Table 1: Comparison of relative importance values for simulation example 1.

| CHANNELS | $\beta$ | $\beta^2$ | $\rho^2$ | $\beta \cdot \rho$ | DA | RW |
|---|---|---|---|---|---|---|
| $x_1$ | 3.0 | 0.165 | 0.027 | 0.079 | 0.128 (0.036) | 0.126 (0.049) |
| $x_2$ | -4.5 | 0.372 | 0.432 | 0.474 | 0.342 (0.041) | 0.344 (0.058) |
| $x_3$ | -0.5 | 0.005 | 0.108 | 0.026 | 0.054 (0.027) | 0.069 (0.033) |
| $x_4$ | 3.0 | 0.165 | 0.000 | 0.000 | 0.091 (0.021) | 0.086 (0.027) |
| $x_5$ | -4.0 | 0.294 | 0.432 | 0.421 | 0.354 (0.056) | 0.374 (0.081) |

Additionally, Table 1 reports the standard derivations of the importance values by DA and RW over 30 replicates (shown in parentheses). As discussed in the literature, DA has known asymptotic properties (Budescu, 1993; Grömping, 2007). I.e., as we increase the sample size of dataset, the DA estimators converge to theoretical DA values. On the other hand, RW has some convergent validity properties, and it generally produces very similar results to DA in many cases.

Therefore, researchers tend to conclude that they are measuring the same quantities (Johnson and Lebreton, 2004). Our numerical results show that these two methods consistently suggest very close relative importance values, while RW estimators generally have higher variability than DA.

### 4.2 Simulation Example 2: Relative Importance Methods for Highly Correlated Data

We consider a simulated example with higher correlations among variables, and show DA and RW are able to take this into account in calculating attribution values while traditional methods fail to do so. To generate a more extreme simulated dataset, we modify the correlation matrix in previous example to $Cor(x_{j_1}, x_{j_2}) = 0.8$ (i.e., all pairs of variables are equally highly correlated). This simulation example is replicated 30 times as well.

Table 2 reports the theoretical values of $\beta^2, \rho^2, \beta \cdot \rho$ and means & standard derivations of relative importance values over 30 replicates. It can be seen that both DA and RW suggest consistent importance values with low variability. Because of the high correlations among them, there is no enough evidence to clearly tell which variable is significantly more important than another just by looking at the coefficients. In other words, DA and RW can account for model interrelatedness. This is not true for $\boldsymbol{\beta^2}$ and $\boldsymbol{\rho^2}$. Lastly, we note that $\boldsymbol{\beta \cdot \rho}$ is not an effective measure, as negative values have been produced, which is hard to interpret in practice.

Table 2: Comparison of relative importance values for simulation example 2.

| CHANNELS | $\beta$ | $\beta^2$ | $\rho^2$ | $\beta \cdot \rho$ | DA | RW |
|---|---|---|---|---|---|---|
| $x_1$ | 3.0 | 0.165 | 0.096 | -0.298 | 0.110 (0.011) | 0.115 (0.018) |
| $x_2$ | -4.5 | 0.372 | 0.322 | 0.820 | 0.341 (0.029) | 0.334 (0.034) |
| $x_3$ | -0.5 | 0.005 | 0.185 | 0.069 | 0.090 (0.015) | 0.140 (0.023) |
| $x_4$ | 3.0 | 0.165 | 0.096 | -0.298 | 0.111 (0.014) | 0.111 (0.013) |
| $x_5$ | -4.0 | 0.294 | 0.302 | 0.707 | 0.299 (0.028) | 0.300 (0.037) |

### 4.3 Simulation Example 3: Relative Importance Methods: Linear & Additive Models

This simulation example compares the performance of linear and additive models. We use a simulated dataset generated from a sophisticated non-linear model as following,

$$y = \sum_{j=1}^{5} f_j(x_j) + \varepsilon, \tag{15}$$

with $f_1(x) = x(1-x)$, $f_2(x) = 2\log(\max(x,1))$, $f_3(x) = 1 - \exp(-x)$, $f_4(x) = 2x^{2/5}$ and $f_5(x) = x$. Functions $f_2(\cdot) - f_5(\cdot)$ are commonly used as saturating and non-saturating transformation in marketing practice, while $f_1(\cdot)$ represents an arbitrary nonlinear function. $\varepsilon$ is the error term following standard normal distribution. We first generate 1000 training data points in which variables $x_j$'s are generated from the standard normal distribution with correlation matrix given by $Cor(x_{j_1}, x_{j_2}) = (1/2)^{|j_1 - j_2|}$. We apply DA and RW with linear and additive models to this training set. We apply 5-fold cross validation to select the number of TPS internal knots in additive model. To assess the model performance, we generate a test dataset from the same model. It consists of one thousand observations. Test error is measured using root-mean-squared error (RMSE). More specifically,

$$RMSE = \sqrt{\frac{1}{1000} \sum_{i=1}^{1000} (\hat{y}_i - y_i)^2}, \tag{16}$$

where $y_i, i = 1, \ldots, 1000$ are the true values of dependent variable in the test dataset, and $\hat{y}_i, i = 1, \ldots 1000$ are predicted values. We replicate this training/testing simulation 30 times.

Table 3: Comparison of linear and additive models.

| CHANNELS | LINEAR MODEL | | ADDITIVE MODEL | |
|---|---|---|---|---|
| | DA | RW | DA | RW |
| $x_1$ | 0.191 (0.025) | 0.199 (0.026) | 0.289 (0.028) | 0.279 (0.028) |
| $x_2$ | 0.136 (0.013) | 0.130 (0.013) | 0.102 (0.010) | 0.116 (0.009) |
| $x_3$ | 0.402 (0.031) | 0.400 (0.032) | 0.396 (0.037) | 0.384 (0.041) |
| $x_4$ | 0.105 (0.014) | 0.099 (0.014) | 0.099 (0.016) | 0.102 (0.011) |
| $x_5$ | 0.166 (0.019) | 0.173 (0.020) | 0.114 (0.015) | 0.118 (0.015) |
| $R^2$ | 0.614 (0.045) | | 0.959 (0.039) | |
| RMSE | 0.646 (0.025) | | 0.174 (0.035) | |

Table 3 presents the relative importance of the variables (channels), $R^2$ of the fitted models and the RMSE. Similarly, we report the means and standard derivations of them over 30 replicates. As reported, linear and additive models suggest similarly ranked yet somewhat different relative importance values. We observe that the additive model provides a better fit for the data and higher predictive accuracy. This is because, with less parametric restriction, additive model is more flexible by allowing the data to choose the proper functional components. With higher $R^2$ and predictive ability, the importance values suggested by the additive model are more trustworthy. This is also most likely to be true in analyzing any real data which generally has more sophisticated underlying relationship than our simulation example.

## 4.4 Real Data: Group Analysis

In this section, we apply all aforementioned methods to a real dataset. The data we use is event-level records of online users with revenue generating tracking identifiers from a 3-month campaign, and the associated users' exposures to 18 advertising channels. These channels are grouped into three categories: Publishers, DSPs, Paid Search. We pre-process the raw data by matching the unique channel ID's and grouping them according to the users so that revenue and exposures from each user can be tracked. As a result, we have a total of 153,891 revenue-generating observations. Table 4 presents a detailed description of the dataset. Note that total revenue for each category in the table should not be confused with attribution value. It contains revenue from all users who have visited this category. However, these users may have visited other category as well so that their revenue may be counted more than once.

Table 4: Data Summary

|  | PUBLISHERS | DSP'S | PAID SEARCH | TOTAL |
|---|---|---|---|---|
| Total Revenue (M$) | 2.2 | 6.2 | 6.8 | 14 |
| Impressions | $1.0 * 10^5$ | $8.5 * 10^5$ | $9.7 * 10^4$ | $1.1 * 10^6$ |
| Average Revenue per Impression ($) | 21 | 7.2 | 69 | 14 |

An important practical issue with real data is that some channels may have negative coefficients in linear models. Since the proposed approaches are based on decomposition of $R^2$ with guaranteed non-negativity, they always produce non-negative attribution values. However, the negative coefficients are hard to interpret and use in media investment planning practice, especially in calculating Return on Investment (ROI) for each individual channel. Therefore, we further produce hybrid attribution values in addition to the original results. More precisely, we filter the channels by only keeping those with positive regression coefficients and then renormalize their attribution values, while we assign no attribution to the rest of the channels.

We utilize the proposed methods: (i) DA with linear models (DALM), (ii) DA with additive models (DAAM), (iii) RW with linear models (RWLM) and (iv) RW with additive models (RWAM), to calculate the attribution values of the aforementioned advertising channel groups. We report the resulting relative attribution values, running times and the coefficient of determination ($R^2$) in Table 5.

Table 5: Relative attribution values for group analysis

| CHANNEL GROUPS | $\beta$ | DA | | RW | |
|---|---|---|---|---|---|
| | | Linear (LM) | Additive (AM) | Linear (LM) | Additive (AM) |
| Publishers | 5.4% | 0.62% | 2.1% | 0.64% | 2.5% |
| DSPs | 38% | 29% | 33% | 29% | 35% |
| Paid Search | 57% | 71% | 65% | 71% | 62% |
| Running Time | 0.11s | 0.35s | 3.5mins | 0.11s | 2.7mins |
| $R^2$ | - | 0.06 | 0.15 | 0.06 | 0.15 |

The attribution values are also presented in Figure 1. As we observe, all the proposed methods consistently suggest that paid search has the largest attribution value which is at least 62%, followed by DSPs (≥29%) and publishers (< 3%).

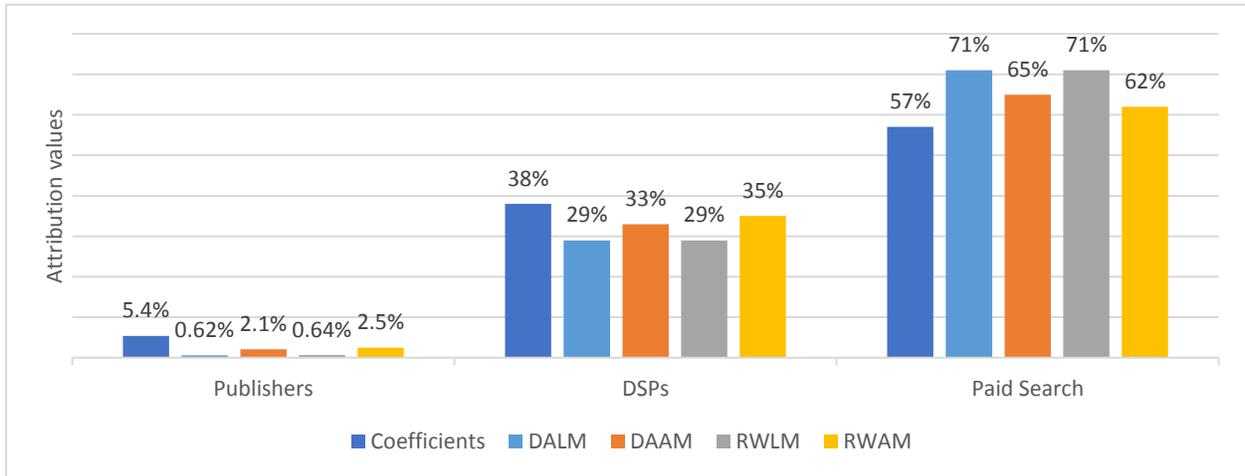

Figure 1: Attribution values for group analysis

By comparing the attribution values obtained from the same regression models (i.e. DALM vs RWLM and DAAM vs RWAM), we notice that attribution values suggested by the same model structure are very close (Tonidandel and LeBreton, 2011).

We note obvious differences in the attribution values as well as goodness-of-fit when comparing the results across the models. Specifically, additive models tend to put more weights on DSPs which is around 34% and publishers which is around 2.3%, while reducing the attribution value of paid search to about 64%. Even though justifying the results for real data is generally difficult, an additive model way of decomposing attributions naturally describes a closer fit to the true underlying relationship and thus can be more likely to be trustworthy.

As we have stated earlier, DA suffers from heavier computational burden than RW does. This could be even worse as the number of channels increases. Therefore DA is less preferable for large-scale applications.

## 4.5   Real Data: All-channel Analysis

Instead of attributing revenue to the channel groups, we run an all-channel analysis for the same dataset in the previous section. Specifically, we apply RWLM and RWAM to analyze how revenue should be attributed to 18 advertising channels. As DA always produces very close results to those of RW, but with prohibitively higher computational cost (and exponentially higher so with the increased number of independent variables to be evaluated), we choose not to utilize DALM and DAAM in this section. The channels are named as P1, P2, etc. for those belonging to publishers, D1 for a channel from DSPs, and S1 and S2 for channels belonging to paid search.

In Figure 2, we plot the attribution values calculated by RWLM and RWAM, and the corresponding data is shown in Table 6. Both models consistently report S1, S2 and D1 as the top three attributed channels. Nevertheless, we also see some inconsistencies in attribution values. Specifically, RWLM assigns a significantly higher attribution value to channel S1 in comparison to RWAM (45% vs. 35%). In contrast, the attribution values of D1 and S2 by RWLM and RWAM are relatively close. Attributions to these three channels are closer to each other under RWAM than RWLM.

An important observation is that some channels (P10, P13, P14 and P15) have negative regression coefficients. In fact, those channels are all negatively correlated with the dependent variable and thus may potentially make negative contribution to the revenue. It is also possibly because that P10, P13, P14 and P15 are relatively highly correlated to other channels so that positive marginal contribution of them may be taken up by others. Unsurprisingly, as shown in Table 6, the raw attributions values of those channels are negligible. As mentioned previously, we choose to further produce hybrid results by disregarding these four channels and renormalizing the attribution values for the rest.

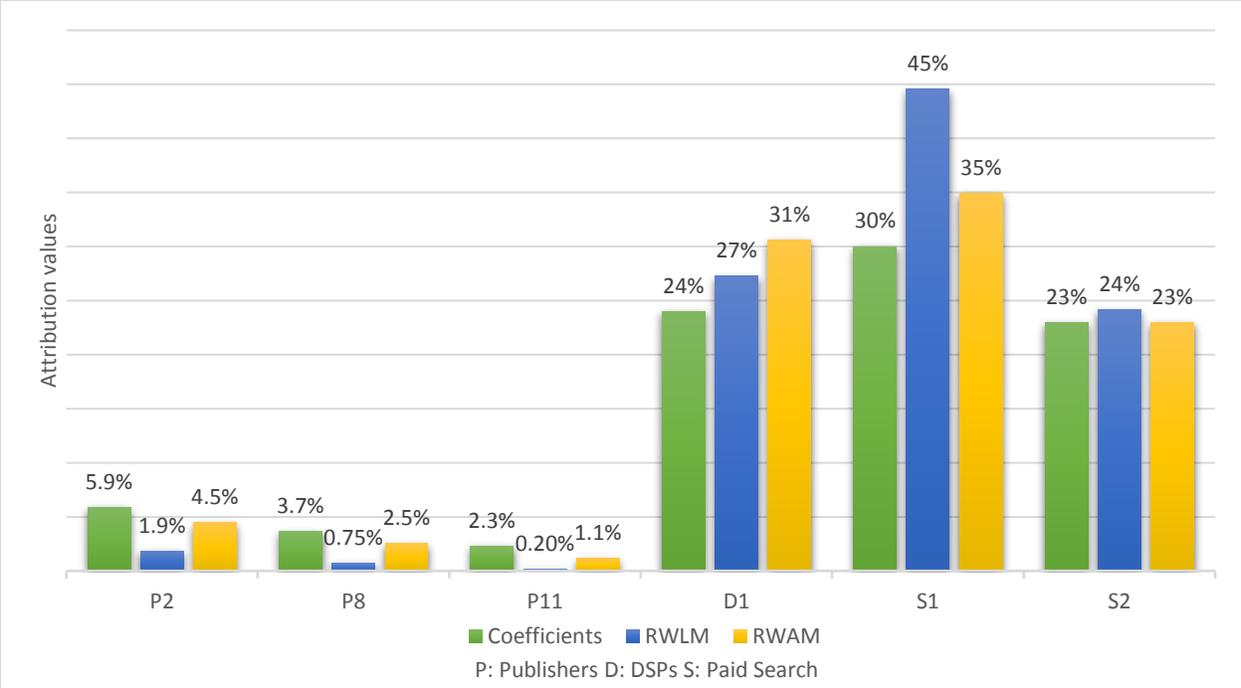

Figure 2. Attribution values for all channel analysis. Note that we omit channels with attribution values less than 1% due to space limitation.

Table 6: Attribution values for all-channel analysis

| CHANNEL | β | RWLM | | RWAM | |
|---|---|---|---|---|---|
| | | Raw | Hybrid | Raw | Hybrid |
| P1 | 0.21% | 0.010% | 0.010% | 0.17% | 0.17% |
| P2 | 5.9% | 1.8% | 1.9% | 4.4% | 4.5% |
| P3 | 0.35% | 0.030% | 0.030% | 0.41% | 0.42% |
| P4 | 2.3% | 0.26% | 0.26% | 0.44% | 0.45% |
| P5 | 0.49% | 0.010% | 0.010% | 0.091% | 0.091% |
| P6 | 2.1% | 0.20% | 0.20% | 0.52% | 0.53% |
| P7 | 2.1% | 0.24% | 0.24% | 0.30% | 0.30% |
| P8 | 3.7% | 0.74% | 0.75% | 2.5% | 2.6% |
| P9 | 2.3% | 0.29% | 0.29% | 0.60% | 0.61% |
| p10 | 0% | 0.86% | 0% | 0.66% | 0% |
| P11 | 2.3% | 0.20% | 0.20% | 1.1% | 1.1% |
| P12 | 1.7% | 0.16% | 0.16% | 0.69% | 0.70% |
| P13 | 0% | 0.15% | 0% | 0.14% | 0% |
| P14 | 0% | 0.051% | 0% | 0.50% | 0% |
| P15 | 0% | 0.070% | 0% | 0.18% | 0% |
| D1 | 24% | 27% | 27% | 30% | 31% |
| S1 | 30% | 44% | 45% | 34% | 35% |
| S2 | 23% | 24% | 24% | 23% | 23% |

# 5   Conclusion and Discussion

In this paper, we have developed several algorithmic methods for allocating online advertising exposure to multiple channels through attribution models. The proposed approaches bridge the gap between attribution modeling in marketing and the related methodologies developed in statistics. Moreover, they provide a more accurate overview of our multi-touch attribution problem by capturing the co-occurrences and other interactions among advertising channels.

We implement relative importance methods, including dominance analysis and relative weight analysis, with extension to semi-parametric (additive) models. Our work can be potentially extended to a much richer pool of regression models that marketing researchers may find useful. More specifically, dominance analysis can be readily applied to other types of regression models, such as partially linear additive models (Liang et al., 2008; Ma and Yang, 2011; Zhao and Lian, 2014), varying-coefficient models (Fan and Zhang, 2008; Hastie and Tibshirani, 1993) and others. While relative weight analysis generally works well for linear models: those are linear in coefficients, such as marketing mix models (Ansari et al., 1995; Naik et al., 2005; Ramaswamy et al., 1993), vector autoregressive models (Hamilton, 1994) and others.

We note that the proposed approaches come with some inherent advantages and limitations in terms of scalability. Time performance of dominance analysis quickly deteriorates with the growing number of channels. However, the method is easily scalable to an immense user set as long as the number of channels are kept constant. On the other hand, relative weight analysis is easily scalable to multiple channels. However, the computational burden increases with the number of users (volume) increases. In principle, dominance analysis is feasible for analysis with larger volume and less channels, while relative weight analysis is practical for analysis with smaller volume and more channels.

One common strength the two proposed data-driven revenue-based attribution models share, apart from accurately accounting for interrelatedness as inferred through attribution models, is consistency and robustness across resamples. Through several use cases, we have consistently reported relatively close attribution values. Therefore the proposed approaches can scale both for the "big data" and the "omni-channel" analysis. More accurate and consistent attribution values will also help advertising practitioners make smarter decisions in advertising strategy and planning.

# References

Abhishek, V., Fader, P., Hosanagar, K., 2012. Media Exposure through the Funnel: A Model of Multi-Stage Attribution (SSRN Scholarly Paper No. ID 2158421). Social Science Research Network, Rochester, NY.

Anderl, E., Becker, I., v. Wangenheim, F., Schumann, J.H., 2013. Putting Attribution to Work: A Graph-Based Framework for Attribution Modeling in Managerial Practice. SSRN Electron. J. doi:10.2139/ssrn.2343077

Ansari, A., Bawa, K., Ghosh, A., 1995. A nested logit model of brand choice incorporating variety-seeking and marketing-mix variables. Mark. Lett. 6, 199–210. doi:10.1007/BF00995111

Azen, R., Budescu, D.V., 2003. The dominance analysis approach for comparing predictors in multiple regression. Psychol. Methods 8, 129–148. doi:10.1037/1082-989X.8.2.129

Budescu, D.V., 1993. Dominance analysis: A new approach to the problem of relative importance of predictors in multiple regression. Psychol. Bull. 114, 542–551. doi:10.1037/0033-2909.114.3.542

Cain, P.M., 2010. Marketing Mix Modelling and Return on Investment, in: Kitchen, P.J. (Ed.), Integrated Brand Marketing and Measuring Returns. Palgrave Macmillan UK, London, pp. 94–130.

Dalessandro, B., Perlich, C., Stitelman, O., Provost, F., 2012. Causally Motivated Attribution for Online Advertising, in: Proceedings of the Sixth International Workshop on Data Mining for Online Advertising and Internet Economy, ADKDD '12. ACM, New York, NY, USA, p. 7:1–7:9. doi:10.1145/2351356.2351363

Fan, J., Zhang, W., 2008. Statistical methods with varying coefficient models. Stat. Interface 1, 179.

Friedman, J., Hastie, T., Tibshirani, R., 2001. The elements of statistical learning. Springer series in statistics Springer, Berlin.

Green, P.E., Tull, D.S., 1974. Research for marketing decisions. Prentice-Hall.

Grömping, U., 2007. Estimators of Relative Importance in Linear Regression Based on Variance Decomposition. Am. Stat. 61, 139–147.

Hamilton, J.D., 1994. Time series analysis. Princeton university press Princeton.

Hastie, T., Tibshirani, R., 1993. Varying-Coefficient Models. J. R. Stat. Soc. Ser. B Methodol. 55, 757–796.

Hastie, T.J., Tibshirani, R.J., 1990. Generalized Additive Models. CRC Press.

Hoffman, P.J., 1960. The paramorphic representation of clinical judgment. Psychol. Bull. 57, 116–131.

IAB_Internet_Advertising_Revenue_FY_2014.pdf, n.d.

James, G., Witten, D., Hastie, T., Tibshirani, R., 2013. An Introduction to Statistical Learning, Springer Texts in Statistics. Springer New York, New York, NY.

Johnson, J.W., 2000. A Heuristic Method for Estimating the Relative Weight of Predictor Variables in Multiple Regression. Multivar. Behav. Res. 35, 1–19. doi:10.1207/S15327906MBR3501_1

Johnson, J.W., Lebreton, J.M., 2004. History and Use of Relative Importance Indices in Organizational Research. Organ. Res. Methods 7, 238–257. doi:10.1177/1094428104266510

Li, H. (Alice), Kannan, P. k., 2013. Attributing Conversions in a Multichannel Online Marketing Environment: An Empirical Model and a Field Experiment. J. Mark. Res. 51, 40–56. doi:10.1509/jmr.13.0050

Liang, H., Thurston, S.W., Ruppert, D., Apanasovich, T., Hauser, R., 2008. Additive Partial Linear Models with Measurement Errors. Biometrika 95, 667–678.

Ma, S., Yang, L., 2011. Spline-backfitted kernel smoothing of partially linear additive model. J. Stat. Plan. Inference 141, 204–219. doi:10.1016/j.jspi.2010.05.028

Murphy, K.P., 2012. Machine learning: a probabilistic perspective, Adaptive computation and machine learning series. MIT Press, Cambridge, MA.

Naik, P.A., Raman, K., Winer, R.S., 2005. Planning Marketing-Mix Strategies in the Presence of Interaction Effects. Mark. Sci. 24, 25–34. doi:10.1287/mksc.1040.0083

Osborne, M.J., Rubinstein, A., 1994. A course in game theory. MIT press.

R., J., de Boor, C., 1980. A Practical Guide to Splines. Math. Comput. 34, 325. doi:10.2307/2006241

Ramaswamy, V., Desarbo, W.S., Reibstein, D.J., Robinson, W.T., 1993. An Empirical Pooling Approach for Estimating Marketing Mix Elasticities with PIMS Data. Mark. Sci. 12, 103–124. doi:10.1287/mksc.12.1.103

Shao, X., Li, L., 2011. Data-driven Multi-touch Attribution Models, in: Proceedings of the 17th ACM SIGKDD International Conference on Knowledge Discovery and Data Mining, KDD '11. ACM, New York, NY, USA, pp. 258–264. doi:10.1145/2020408.2020453

Shapley, L.S., 1952. A value for n-person games. DTIC Document.

Tonidandel, S., LeBreton, J.M., 2011. Relative Importance Analysis: A Useful Supplement to Regression Analysis. J. Bus. Psychol. 26, 1–9. doi:10.1007/s10869-010-9204-3

Tucker, C., 2012. Implications of Improved Attribution and Measurability for Antitrust and Privacy in Online Advertising Markets, The. Geo Mason Rev 20, 1025.

Wiesel, T., Pauwels, K., Arts, J., 2011. Practice Prize Paper —Marketing's Profit Impact: Quantifying Online and Off-line Funnel Progression. Mark. Sci. 30, 604–611. doi:10.1287/mksc.1100.0612

Yadagiri, M.M., Saini, S.K., Sinha, R., 2015. A Non-parametric Approach to the Multi-channel Attribution Problem, in: Wang, J., Cellary, W., Wang, D., Wang, H., Chen, S.-C., Li, T., Zhang, Y. (Eds.), Web Information Systems Engineering – WISE 2015, Lecture Notes in Computer Science. Springer International Publishing, pp. 338–352.

Zhao, K., Lian, H., 2014. Variational inferences for partially linear additive models with variable selection. Comput. Stat. Data Anal. 80, 223–239.